\documentclass{article}
\usepackage{spconf,amsmath,amssymb,graphicx,hyperref}
\usepackage{multirow}
\usepackage{adjustbox}
\usepackage{booktabs}

\newcommand{\msd}[2]{%
  \ensuremath{#1{\scriptscriptstyle\,\pm\,#2}}%
}

\newcommand{\bmsd}[2]{%
  \ensuremath{\mathbf{#1}{\scriptscriptstyle\,\pm\,#2}}%
}

\title{Segmented vs. Contextual Phoneme Encoding for Cross-Language Audio Deepfake Detection}

\name{
\begin{tabular}{c}
Vamshi Nallaguntla$^{1}$ \quad
Aishwarya Fursule$^{1}$ \quad
Yuan Tseng$^{2}$ \\[2pt]
David Harwath$^{2}$
\quad
Anderson Avila$^{3}$ \quad
Shruti Kshirsagar$^{1}$ 
\end{tabular}
}

\address{
\begin{tabular}{c}
$^{1}$Wichita State University, USA \qquad
$^{2}$The University of Texas at Austin, USA \\[2pt]
$^{3}$Institut national de la recherche scientifique (INRS-EMT), Canada
\end{tabular}
}

\begin{document}
%
\maketitle

\begin{abstract}
Synthetic speech poses growing security risks, yet most audio deepfake
detectors make utterance-level decisions that reveal little about where
spoofing cues lie or whether they transfer to unseen languages. We
investigate phonetic-category representations for cross-language
deepfake detection. We group language-specific phones into eight broad
categories and compare two systems: segmented phoneme encoding, which
encodes each cropped phone independently, and contextual phoneme
encoding, which encodes the full utterance before pooling within phone
boundaries. Using the MLAAD dataset, we train on three languages and
evaluate on five unseen languages. Contextual encoding substantially
outperforms segmented encoding across all fusion strategies and target
languages. Monophthongs and silence provide the strongest standalone
cues under contextual encoding. Silence-removal experiments show that
segmented encoding depends heavily on silence, whereas contextual
encoding remains largely robust. These findings highlight the importance of phonetic context
for analyzing synthetic-speech cues across languages.
\end{abstract}


\section{Introduction}
\label{sec:intro}

Modern text-to-speech and voice-conversion systems can now produce
convincing speech in many languages. This creates real security risks,
including voice-based fraud, impersonation, and attacks on speaker
verification systems, which makes reliable synthetic speech detection
an essential line of defense. Detection methods have evolved from
handcrafted acoustic features to deep models trained on raw waveforms
and learned representations \cite{delgado2021asvspoof,zhang2025audio}.
Many recent systems build on self-supervised learning (SSL) encoders
such as WavLM \cite{chen2022wavlm} and wav2vec~2.0
\cite{baevski2020wav2vec}, and related work has begun to examine the
biases these detectors inherit from their training data
\cite{fursule2026you}. Although these detectors perform well, they
typically make a single decision for the whole utterance. As a result,
they reveal little about \emph{where} in the speech signal the evidence
of synthesis lies, or whether that evidence holds in languages the
detector has never seen.  To answer these questions, several studies have moved below the
utterance level. Yang et al.~\cite{yang2025forensic} showed that
phoneme-level features provide complementary forensic information.
Temmar et al.~\cite{temmar2025phonetic} and Nallaguntla et
al.~\cite{nallaguntla2026phonemedf,nallaguntla2026phoneme} found that
differences between real and synthetic speech vary from one phoneme to
another. Building on this idea, Zhang et al.~\cite{zhang2025phoneme}
used phoneme-level discrepancies for cross-language detection, while
Tseng et al.~\cite{tseng2026spade} introduced the SPADE dataset and
showed that multilingual partial deepfake detection remains
challenging. Sivaraman et al.~\cite{sivaraman2025investigating} further
analyzed speech regions to make detection more explainable. Despite this
progress, two problems remain. First, each language has its own phone
inventory, so findings for one language do not transfer easily to
another. Second, most phone-level studies cut out each phone and encode
it in isolation, discarding the surrounding context that SSL encoders
are designed to exploit. In this work, we address both problems. To make phones comparable
across languages, we group them into eight broad categories: silence, fricatives, affricates, plosives, vibrants,
nasals, approximants, and monophthongs. 
To study the role of context, we compare two systems.
System 1: Segmented Phoneme Encoding (SPE) and system 2: Contextual Phoneme Encoding (CPE). Our main contributions are as follows: (1) We compare the Segmented Phoneme Encoding and Contextual Phoneme Encoding systems with four fusion methods and evaluated on all five unseen
    languages. (2) We train a separate detector for each phonetic category and evaluate it on all five languages with both systems. (3) We conducted ablation study on silence across
    six SSL encoders to test whether our findings hold beyond a single model.

The remainder of this paper is organized as follows.
Section~\ref{sec:methodology} describes the proposed methodology including dataset, the two systems, encoder, classifier and the fusion and evaluation strategies.
Section~\ref{sec:results} presents and discusses the experimental
results. Finally, Section~\ref{sec:conclusion} concludes the paper and
outlines directions for future work.

\begin{figure*}[t]
    \centering
    \includegraphics[width=0.95\textwidth]{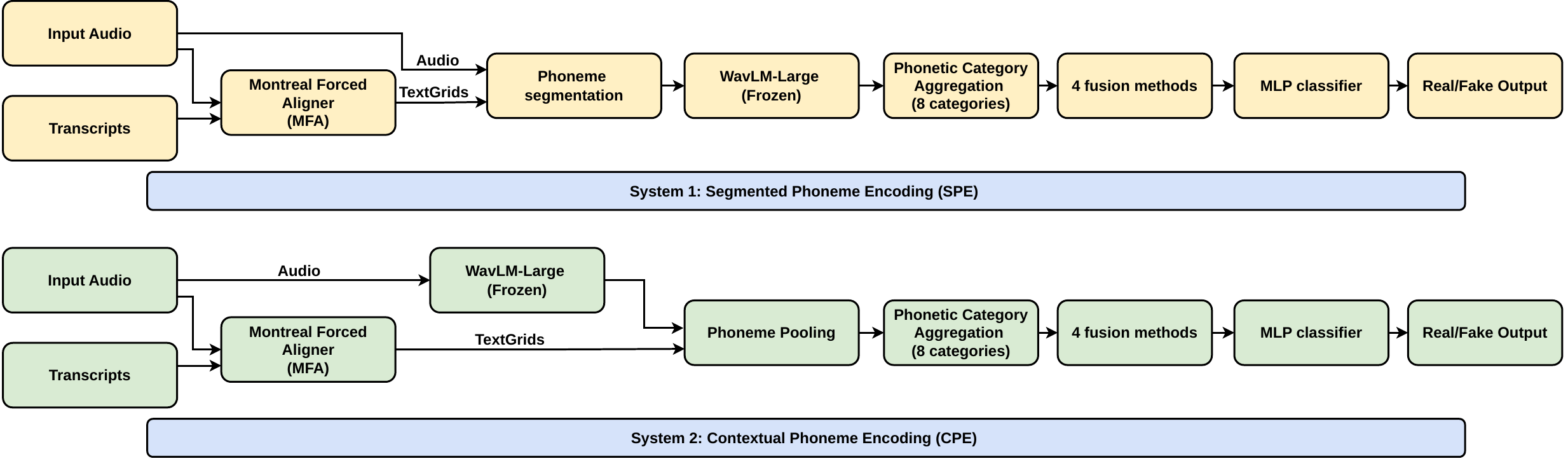}
    \caption{Overview of the two systems}
    \label{fig:proposed_framework}
\end{figure*}

\section{Proposed Methodology}
\label{sec:methodology}
In this section, we present the proposed methodology. Figure~\ref{fig:proposed_framework} illustrates
the two encoding systems evaluated in this study.

\noindent\textbf{System 1: Segmented Phoneme
Encoding (SPE).}
The input waveform is segmented into individual
phone waveforms using MFA-derived boundaries.
Each segment is independently processed by
the frozen SSL encoder, and its frame-level
features are mean-pooled to obtain a phone
representation. These representations are
aggregated into phonetic categories, combined
using the four fusion strategies, and passed
to the binary classifier.

\noindent\textbf{System 2: Contextual Phoneme
Encoding (CPE).}
The complete utterance is first processed
by the frozen SSL encoder. MFA-derived phone
boundaries are then used to mean-pool the
corresponding encoder frames into phone-level
representations. These are aggregated into
phonetic categories, fused, and passed to
the binary classifier. Unlike SPE, CPE
retains surrounding acoustic context in
each phone representation.

\subsection{Dataset}
\label{sec:dataset}

We use the Multilingual Audio Anti-Spoofing Dataset (MLAAD) \cite{muller2024mlaad}, which includes bona fide speech from M-AILABS \cite{mailabs2023dataset}. We select German (DE), English (EN), and Spanish (ES) as source languages and reserve French (FR), Italian (IT), Polish (PL), Russian (RU), and Ukrainian (UK) for unseen-language evaluation. We split the source-language data into training (85\%) and development (15\%) sets, stratifying by language and class. This process yields 23,088 training utterances and 4,078 development utterances. We use 36,943 utterances from the five unseen languages to evaluate the models' cross-language generalization capabilities.

\subsection{Montreal Forced Alignment}
\label{sec:alignment}

We obtain phone boundaries using the Montreal Forced Aligner (MFA)
\cite{mcauliffe2017montreal} with language-specific acoustic models
and pronunciation dictionaries. We exclude utterances with missing
or invalid alignments or with at least one spn interval, and we
retain non-speech intervals as silence. Of 100,771 initial utterances, 99,114 have usable audio--TextGrid
pairs. Excluding the 35,005 utterances that contain spn leaves
64,109 utterances (63.6\%).

\subsection{Speech Encoder Representations}
\label{sec:encoder}

We employ WavLM-Large \cite{chen2022wavlm}
as the primary frozen speech encoder.
Its parameters remain fixed throughout
training, allowing us to focus on the
effect of segmented versus contextual
phoneme encoding. To assess encoder
robustness, we additionally evaluate six
frozen SSL checkpoints: HuBERT Base and
Large \cite{hsu2021hubert}, WavLM Base,
Base+, and Large \cite{chen2022wavlm},
and XLS-R 300M \cite{babu2021xls}.

\subsection{Phonetic Category Representation}
\label{sec:phonetic_categories}

We map language-specific phones into eight broad
phonetic categories: silence (SIL), fricative (FRC),
affricate (AFF), plosive (PLO), vibrant (VIB),
nasal (NAS), approximant (APR), and monophthong (MPH).

For an aligned phone $p_i$, let $\mathcal{T}_i$
denote the corresponding encoder frames, which we
obtain either by encoding the segmented phone
independently (SPE) or by selecting its aligned
frames from the full-utterance representation (CPE).
We obtain its representation through mean pooling:
\begin{equation}
\mathbf{z}_i =
\frac{1}{|\mathcal{T}_i|}
\sum_{t\in\mathcal{T}_i}\mathbf{h}_t.
\end{equation}

We subsequently aggregate phones assigned to the
same category:
\begin{equation}
\mathbf{z}_c =
\frac{1}{|\mathcal{P}_c|}
\sum_{i\in\mathcal{P}_c}\mathbf{z}_i,
\end{equation}
where $\mathcal{P}_c$ denotes the set of phones
belonging to category $c$. A binary presence mask
$m_c$ indicates whether category $c$ occurs in
an utterance.

\subsection{Phonetic Category Fusion}
\label{sec:fusion}
Given category representations
$\{\mathbf{z}_c\}_{c=1}^{C}$ and presence
mask $m_c$, we compute the fused representation as
\begin{equation}
\mathbf{z}_{\mathrm{ph}} =
\sum_{c=1}^{C}\alpha_c\mathbf{z}_c,
\end{equation}
where $\alpha_c$ denotes the normalized
category weight. We compare four fusion strategies.

\noindent\textbf{Uniform Fusion.}
Present categories contribute equally:
\begin{equation}
\alpha_c=\frac{m_c}{\sum_j m_j}.
\end{equation}
This provides a parameter-free baseline.

\noindent\textbf{Static Fusion.}
Trainable source-learned scalars $a_c$ determine
global category preferences through a masked
softmax:
\begin{equation}
\alpha_c =
\frac{m_c\exp(a_c)}
{\sum_{j=1}^{C}m_j\exp(a_j)}.
\end{equation}
All utterances share these weights, and absent
categories receive zero weight.

\noindent\textbf{Attention Fusion.}
We compute utterance-dependent category scores as
\begin{equation}
e_c=\mathbf{w}_2^{\top}
\tanh(\mathbf{W}_1\mathbf{z}_c+\mathbf{b}_1)+b_2,
\end{equation}
and normalize them using the masked softmax,
replacing $a_c$ with $e_c$.

\noindent\textbf{PPO Fusion.}
To examine whether policy-based adaptation
benefits category fusion, we investigate
Proximal Policy Optimization (PPO), following
the adaptive-fusion principle of AVATAR
\cite{shabbir2026avatar}.
The policy predicts utterance-dependent weights
over available categories, and we optimize it
using the reward
\begin{equation}
R=\mathbb{I}(\hat{y}=y)
-\lambda D_{\mathrm{TV}}(\boldsymbol{\alpha},
\mathbf{u}), \qquad \lambda=0.5,
\end{equation}
where $\mathbb{I}(\hat{y}=y)$ rewards correct
classification, $\boldsymbol{\alpha}$ is the
predicted weight distribution, $\mathbf{u}$
is uniform over present categories, and
$D_{\mathrm{TV}}$ is the normalized total-variation
distance. This regularization term discourages
excessive weight concentration.

\subsection{Classifier}
\label{sec:classifier}
We pass the fused phonetic-category representation
through a lightweight multilayer perceptron (MLP)
for binary bona fide/spoof classification. The
classifier consists of LayerNorm, a 256-dimensional
linear layer, GELU activation, dropout ($0.3$), and
a linear output layer that produces a single logit. We train the classifier using a weighted binary
cross-entropy loss and AdamW with a learning rate
of $10^{-3}$, weight decay of $10^{-4}$, and batch
size of 256.

\subsection{Evaluation Metrics}
\label{sec:evaluation}

Performance is evaluated using the area under
the receiver operating characteristic curve
(AUC) and equal error rate (EER). AUC measures
the detector's ability to distinguish bona fide
from spoof speech across decision thresholds
and is defined as
\begin{equation}
\mathrm{AUC} =
\int_{0}^{1}\mathrm{TPR}(u)\,du,
\end{equation}
where $u$ denotes the false positive rate
(FPR), and TPR is the true positive rate.

EER is the operating point at which the
false positive rate and false negative rate
(FNR) are equal:
\begin{equation}
\mathrm{EER} =
\mathrm{FPR}(\tau^*) =
\mathrm{FNR}(\tau^*),
\end{equation}
where $\tau^*$ is the decision threshold
at which the two error rates intersect.
\begin{table}[t]
\centering
\caption{Performance comparison of two systems using four fusion methods in terms of EER and AUC.}
\label{tab:main_results}
\begin{adjustbox}{max width=\columnwidth}
\begin{tabular}{lcccc}
\hline
& \multicolumn{2}{c}{System 1}
& \multicolumn{2}{c}{System 2} \\
\cline{2-3}\cline{4-5}
Method & AUC  & EER (\%) 
       & AUC  & EER (\%)  \\

\hline
Uniform
& \msd{.8996}{.0018}
& \msd{17.13}{.23}
& \msd{.9909}{.0003}
& \msd{4.74}{.10} \\

Static
& \bmsd{.9331}{.0022}
& \bmsd{13.32}{.30}
& \msd{.9933}{.0010}
& \msd{4.03}{.32} \\

Attention
& \msd{.9125}{.0023}
& \msd{15.77}{.22}
& \bmsd{.9953}{.0019}
& \bmsd{3.31}{.73} \\

PPO
& \msd{.9081}{.0060}
& \msd{16.26}{.64}
& \msd{.9916}{.0003}
& \msd{4.47}{.11} \\
\hline
\end{tabular}
\end{adjustbox}
\end{table}

\begin{table}[t]
\centering
\caption{Performance comparison of two systems using four fusion methods for five language test set}
\label{tab:language_results}
\setlength{\tabcolsep}{2.5pt}
\renewcommand{\arraystretch}{1.10}
\begin{adjustbox}{max width=\columnwidth}
\begin{tabular}{llccccc}
\hline
Fusion & Enc. & FR & IT & PL & RU & UK \\
\hline

Uniform & System 1
& \msd{11.04}{.42}
& \msd{24.91}{.35}
& \msd{18.72}{.56}
& \msd{9.95}{.33}
& \msd{9.07}{.35} \\

& System 2
& \msd{2.22}{.10}
& \msd{3.86}{.09}
& \msd{4.47}{.15}
& \msd{3.25}{.26}
& \msd{6.77}{.45} \\
\hline

Static & System 1
& \msd{6.68}{.22}
& \msd{22.61}{.26}
& \msd{12.35}{.43}
& \msd{4.44}{.20}
& \msd{7.39}{.62} \\

& System 2
& \msd{1.89}{.20}
& \bmsd{3.29}{.22}
& \bmsd{3.34}{.62}
& \msd{2.42}{.52}
& \msd{5.06}{.84} \\
\hline

Attention & System 1
& \msd{4.11}{.51}
& \msd{27.77}{.45}
& \msd{17.99}{.16}
& \msd{4.80}{.49}
& \msd{7.29}{.37} \\

& System 2
& \bmsd{1.33}{.17}
& \msd{4.06}{.60}
& \msd{3.75}{.60}
& \bmsd{2.37}{.66}
& \bmsd{2.62}{.46} \\
\hline

PPO & System 1
& \msd{9.45}{.30}
& \msd{24.58}{1.48}
& \msd{17.16}{.98}
& \msd{8.62}{.41}
& \msd{8.10}{.76} \\

& System 2
& \msd{2.03}{.24}
& \msd{3.56}{.15}
& \msd{3.77}{.58}
& \msd{2.91}{.19}
& \msd{6.71}{.56} \\
\hline
\end{tabular}
\end{adjustbox}
\vspace{1mm}
\end{table}

\begin{table}[t]
\centering
\caption{Performance comparison of two systems using standalone phonetic categories on five unseen languages in terms of EER (\%).}
\label{tab:category_combined}
\setlength{\tabcolsep}{3pt}
\renewcommand{\arraystretch}{1.12}

\begin{adjustbox}{max width=\columnwidth}
\begin{tabular}{@{}lcccccc@{}}
\hline
Cat. & FR & IT & PL & RU & UK & Pool \\
\hline
\multicolumn{7}{@{}l}{\textit{System 1}} \\
\hline

SIL
& \bmsd{3.63}{.23}
& \bmsd{27.62}{.50}
& \bmsd{18.97}{1.39}
& \bmsd{5.46}{.07}
& \bmsd{7.22}{1.32}
& \bmsd{16.37}{.54} \\

FRC
& \msd{34.41}{.23}
& \msd{41.65}{.21}
& \msd{42.14}{.46}
& \msd{28.21}{.36}
& \msd{33.85}{1.67}
& \msd{37.39}{.27} \\

AFF
& \msd{45.30}{12.82}
& \msd{44.89}{1.38}
& \msd{50.65}{3.22}
& \msd{38.15}{.57}
& \msd{37.66}{.65}
& \msd{44.13}{1.15} \\

PLO
& \msd{32.81}{.53}
& \msd{37.89}{1.24}
& \msd{42.70}{2.58}
& \msd{45.17}{.58}
& \msd{43.89}{1.07}
& \msd{40.52}{.94} \\

VIB
& \msd{43.16}{.28}
& \msd{37.84}{.64}
& \msd{43.88}{1.19}
& \msd{37.51}{.88}
& \msd{46.20}{1.10}
& \msd{40.68}{.28} \\

NAS
& \msd{40.84}{.99}
& \msd{40.47}{.30}
& \msd{48.70}{.81}
& \msd{37.15}{.79}
& \msd{41.00}{2.04}
& \msd{41.55}{.36} \\

APR
& \msd{40.36}{1.18}
& \msd{39.22}{.21}
& \msd{43.40}{.60}
& \msd{37.39}{.65}
& \msd{37.77}{1.29}
& \msd{39.86}{.77} \\

MPH
& \msd{40.14}{.70}
& \msd{33.44}{.97}
& \msd{39.58}{.65}
& \msd{38.48}{.08}
& \msd{23.91}{1.86}
& \msd{37.43}{.24} \\
\hline

\multicolumn{7}{@{}l}{\textit{System 2}} \\
\hline

SIL
& \bmsd{1.23}{.18}
& \msd{7.39}{.46}
& \msd{6.42}{.93}
& \bmsd{3.50}{.31}
& \bmsd{2.87}{.27}
& \msd{5.51}{.30} \\

FRC
& \msd{5.52}{.31}
& \msd{12.27}{.32}
& \msd{11.31}{.55}
& \msd{6.47}{.29}
& \msd{17.07}{.52}
& \msd{10.69}{.14} \\

AFF
& \msd{26.82}{1.21}
& \msd{15.18}{.55}
& \msd{17.59}{.13}
& \msd{13.92}{.24}
& \msd{24.56}{.42}
& \msd{17.49}{.52} \\

PLO
& \msd{5.01}{.36}
& \msd{6.36}{.50}
& \msd{8.41}{.82}
& \msd{5.87}{.18}
& \msd{15.89}{.54}
& \msd{8.99}{.24} \\

VIB
& \msd{10.88}{.28}
& \msd{8.75}{.19}
& \msd{21.49}{1.06}
& \msd{11.27}{.40}
& \msd{25.79}{.75}
& \msd{13.72}{.13} \\

NAS
& \msd{9.21}{.58}
& \msd{9.42}{.21}
& \msd{9.49}{.10}
& \msd{5.70}{.11}
& \msd{15.95}{.23}
& \msd{9.73}{.22} \\

APR
& \msd{7.58}{.14}
& \msd{10.79}{.34}
& \msd{15.37}{.51}
& \msd{8.91}{.40}
& \msd{20.23}{.45}
& \msd{12.43}{.33} \\

MPH
& \msd{3.04}{.19}
& \bmsd{4.18}{.57}
& \bmsd{5.29}{.70}
& \msd{3.73}{.21}
& \msd{7.13}{.60}
& \bmsd{5.18}{.24} \\
\hline
\end{tabular}
\end{adjustbox}
\end{table}

\begin{table}[t]
\centering
\caption{Ablation study with silence removal: performance comparison of two systems in terms of EER (\%).}
\label{tab:silence_analysis}
\setlength{\tabcolsep}{2pt}
\renewcommand{\arraystretch}{1.10}

\begin{adjustbox}{max width=\columnwidth}
\begin{tabular}{@{}lcccccc@{}}
\hline
Analysis & FR & IT & PL & RU & UK & Pool \\
\hline

\multicolumn{7}{l}{\textit{System 1}} \\
\hline

Full Static
& \msd{5.56}{0.53}
& \msd{22.10}{0.18}
& \msd{11.39}{0.94}
& \msd{4.01}{0.09}
& \msd{5.70}{0.06}
& \msd{12.34}{0.14} \\

w/o SIL
& \msd{36.18}{1.17}
& \msd{36.26}{0.52}
& \msd{39.66}{1.04}
& \msd{28.12}{1.88}
& \msd{34.92}{4.76}
& \msd{35.71}{0.61} \\

SIL-only
& \msd{7.91}{0.37}
& \msd{26.61}{0.27}
& \msd{18.63}{2.59}
& \msd{7.62}{1.03}
& \msd{10.97}{1.17}
& \msd{17.48}{0.24} \\
\hline

\multicolumn{7}{l}{\textit{System 2}} \\
\hline

Full Static
& \msd{1.89}{0.20}
& \msd{3.27}{0.22}
& \msd{3.34}{0.62}
& \msd{2.42}{0.52}
& \msd{5.06}{0.84}
& \msd{4.03}{0.32} \\

w/o SIL
& \msd{4.54}{0.06}
& \msd{4.91}{0.26}
& \msd{6.15}{0.46}
& \msd{4.57}{0.23}
& \msd{10.78}{0.19}
& \msd{6.24}{0.49} \\

SIL-only
& \msd{6.71}{1.45}
& \msd{8.08}{1.15}
& \msd{7.43}{1.31}
& \msd{12.85}{2.50}
& \msd{9.78}{2.58}
& \msd{10.68}{1.43} \\
\hline

\end{tabular}
\end{adjustbox}
\end{table}

\begin{table}[t]
\centering
\caption{Performance comparison of two systems across six encoders
in terms of pooled EER (\%).}
\label{tab:encoders}
\setlength{\tabcolsep}{2pt}

\begin{adjustbox}{max width=\columnwidth}
\begin{tabular}{lccccc}
\hline
\textbf{Encoder} &
\textbf{Utterance} &
\multicolumn{2}{c}{\textbf{System 1}} &
\multicolumn{2}{c}{\textbf{System 2}} \\
\cline{3-4}\cline{5-6}
& & \textbf{w SIL} & \textbf{w/o SIL}
& \textbf{w SIL} & \textbf{w/o SIL} \\
\hline

HuBERT-Base
& \msd{5.01}{0.03}
& \bmsd{11.01}{0.09}
& \msd{41.60}{0.80}
& \msd{8.01}{0.19}
& \msd{11.97}{0.46} \\

HuBERT-Large
& \msd{9.12}{0.16}
& \msd{18.31}{0.27}
& \msd{41.75}{0.27}
& \msd{13.61}{0.39}
& \msd{20.44}{0.37} \\

WavLM-Base
& \msd{8.52}{0.28}
& \msd{14.39}{0.45}
& \msd{48.88}{0.48}
& \msd{10.20}{0.41}
& \msd{13.63}{0.55} \\

WavLM-Base+
& \msd{3.92}{0.30}
& \msd{20.66}{0.23}
& \msd{45.00}{0.03}
& \msd{5.32}{0.08}
& \msd{8.97}{0.40} \\

WavLM-Large
& \bmsd{1.59}{0.15}
& \msd{12.39}{0.13}
& \msd{35.72}{0.61}
& \msd{4.03}{0.32}
& \msd{6.23}{0.49} \\

XLS-R 300M
& \msd{2.23}{0.08}
& \msd{11.10}{0.92}
& \bmsd{31.97}{0.33}
& \bmsd{2.41}{0.09}
& \bmsd{2.61}{0.10} \\

\hline
\end{tabular}
\end{adjustbox}
\end{table}

\section{Results and Discussion}
\label{sec:results}
In this section, we discuss the results of five sets of experiments,
each described in detail as follows:\\
We first examine whether preserving utterance context improves
category-based detection on unseen languages, and which fusion strategy
generalizes best. We construct fused representations from the eight
phonetic categories using Uniform, Static, Attention, and PPO fusion for
both System~1 and System~2. Table~\ref{tab:main_results} reports
pooled results, and Table~\ref{tab:language_results} reports
language-wise EER. System~2 outperforms System~1 for every fusion strategy and target
language.
Attention fusion yields the lowest System~2 pooled EER
 but exhibits higher seed-to-seed variability than
Static fusion. The best strategy also depends on the target language:
Attention performs best on FR, RU, and UK, whereas Static performs best
on IT and PL. PPO ($4.47\%$) does not outperform Static or Attention
despite its higher complexity; we attribute this to its regularization
toward uniform category weights, which may limit specialization.

Next, to identify which phonetic categories carry discriminative cues
independently, we train a separate detector for each category under
both System~1 and System~2. Training and evaluation use only
category-present utterances, preventing category presence from acting
as a cue. Table~\ref{tab:category_combined} reports language-wise and
pooled EER. System~2 reduces EER for all eight categories. Monophthongs (MPH) and silence (SIL) achieve
the lowest pooled EERs under System~2
($5.18\%$ and $5.51\%$, respectively),
while plosives and nasals also remain
below $10\%$. Under System~1, silence is the only
competitive category, while all remaining categories
yield pooled EERs between $37\%$ and $45\%$. These results indicate that
surrounding acoustic context accounts for much of the discriminative
information within each category.

Given the strong reliance of System~1 on silence, we quantify the
dependence of each system on this category. We mask the silence category at inference
time in the trained Static fusion models
and also evaluate a silence-only condition. All conditions use the same 36,887 test
utterances containing silence. Table~\ref{tab:silence_analysis} reports
the results. Removing silence increases System~1 pooled EER from $12.34\%$ to
$35.71\%$, a mean paired increase of $23.36$ percentage points
(pp), whereas System~2 EER increases from $4.03\%$ to
$6.24\%$. Silence-only inference yields $17.48\%$
and $10.68\%$ EER for System~1 and System~2, respectively. System~1 therefore depends
substantially more on silence than System~2. However, inference-time
masking may introduce a distribution shift, and System~2 silence
representations may retain information from neighboring speech.

Finally, we explore six other SSL encoder with
Static fusion and silence removal: HuBERT
Base and Large, WavLM Base, Base+, and Large, and XLS-R 300M. We also
report the utterance-only baseline for each encoder.
Table~\ref{tab:encoders} summarizes the results. System~2 outperforms System~1 for all six encoders. XLS-R 300M achieves
the lowest System~2 EER, and WavLM-Large achieves the lowest
utterance-only EER. Silence removal degrades System~1 for
every encoder, yielding EERs of $31.97\%$ to $48.88\%$ (relative
increases of $118\%$ to $278\%$). System~2 is considerably more stable,
with post-removal EERs of $2.61\%$ to $20.44\%$ (relative increases of
$8\%$ to $69\%$). The reduced silence dependence of System~2 thus holds
across encoders.

\section{Conclusion}
\label{sec:conclusion}

In this paper, we investigated phonetic-category representations for
cross-language audio deepfake detection. We compared segmented (System~1) and contextual (System~2) phoneme
encoding for cross-language audio deepfake detection. Trained on three
MLAAD languages and tested on five unseen ones, System~2 outperformed
System~1 across all fusion strategies, categories, and languages, with
monophthongs and silence as the strongest cues. System~1 relies heavily
on silence, whereas System~2 remains robust to its removal across six
SSL encoders, highlighting the importance of phonetic context. Future
work will extend this analysis to more datasets and to interpretable
detection.

\vfill\pagebreak

\bibliographystyle{IEEEbib}
\bibliography{strings,refs}

\end{document}